\documentclass[12pt]{article}
\usepackage{epsfig}
\usepackage{graphicx}

\def\beq{\begin{equation}}
\def\eeq#1{\label{#1}\end{equation}}
\def\eeqn{\end{equation}}

\def\beqa{\begin{eqnarray}}
\def\eeqa#1{\label{#1}\end{eqnarray}}
\def\eeqan{\end{eqnarray}}

\let\bar=\overbar

\def\Dslash{\not{\hbox{\kern-4pt $D$}}}
\def\dslash{\not{\hbox{\kern-2pt $\del$}}}

\def\msb{{\bar{\ssstyle M \kern -1pt S}}}

\usepackage{fancyhdr,graphicx}
\def\Title#1{\begin{center} {\Large {\bf #1} } \end{center}}

\newcommand{\lsim}{\stackrel{\scriptstyle <}{\phantom{}_{\sim}}}
\newcommand{\gsim}{\stackrel{\scriptstyle >}{\phantom{}_{\sim}}}

\begin{document}

\Title{Proton gaps and cooling of neutron stars with a stiff hadronic EoS}

\bigskip\bigskip


\begin{raggedright}

{\it
Hovik Grigorian$^{1,2}$,~~Dmitry N. Voskresensky$^{3}$~~and David Blaschke$^{4,5}$\\
\bigskip
$^{1}$Laboratory for Information Technologies,
Joint Institute for Nuclear Reseaarch,
Joliot-Curie Street 6,
141980 Dubna,
Russia\\
\bigskip
$^{2}$Department of Physics,
Yerevan State University,
Alek Manukyan Street 1,
Yerevan 0025,
Armenia\\
%
\bigskip
$^{3}$National Research Nuclear University (MEPhI),
Kashirskoe Shosse 31,
115409 Moscow, Russia\\
\bigskip
$^{4}$Bogoliubov Laboratory for Theoretical Physics,
Joint Institute for Nuclear Research,
Joliot-Curie Street 6,
141980 Dubna,
Russia\\
\bigskip
$^{5}$Institute for Theoretical Physics,
University of Wroclaw,
Max Born Place 9,
50-204 Wroclaw,
Poland\\
}

\end{raggedright}

\vskip 5mm \centerline{\bf Abstract} The recent measurements of
the masses of the pulsar J00737-3039B and of the companion
J1756-2251 and pulsars PSR J1614-2230, PSR J0348-0432  demonstrate
the existence of compact stars with masses in a broad range from 1.2
to 2 $M_\odot$. To fulfill the  constraint $M_{\rm max}>2M_{\odot}$
and to demonstrate  the possibility of cooling scenarios for
purely hadronic and further for hybrid stars we exploit the stiff
DD2 hadronic equation of state producing a maximum neutron star mass
$M\simeq 2.43 M_{\odot}$. We show that the "nuclear medium
cooling" scenario for neutron stars comfortably explains the
whole set of cooling curves just  by a variation of the star masses
without the necessity for the occurrence of the direct Urca reaction. To
describe the cooling data with the very stiff DD2 equation of state we
select a proton gap profile from those exploited in the
literature and allow for a variation of the effective pion gap
controlling the efficiency of the medium modified Urca process.
Fast cooling of young neutron stars like it is seen in the data for Cas A
is explained with the DD2  equation of state when the following conditions
are provided: the presence of an efficient medium modified Urca process,
and  a large proton gap at densities $n\lsim
2n_0$ vanishing for $n\gsim (2.5 - 3) n_0$, where $n_0$ is the
saturation nuclear density. \vskip 10mm

\section{\label{sec:intro}Introduction}

Experimental data on surface temperatures of neutron stars (NSs)
provide us with information about the neutrino emissivities of
various processes, depending on the density behaviour of the $NN$
interaction amplitude, values and density profiles of the proton
and neutron paring gaps, the heat transport and the equation of
state (EoS) of NS matter.
Recently the situation has been improved with the observation of a
segment of the cooling curve for the young  NS in the remnant of the
historical supernova Cassiopeia A (Cas A) \cite{Tananbaum:1999kx,Ashworth:1980vn},
with known age ($\simeq 330$ yr), for which the temperature and the rate of cooling
have been followed over the past 13 years since its discovery
\cite{Ho:2009fk,Heinke:2010xy,Page:2010aw,Yakovlev:2010,Shternin:2010qi,Elshamouty:2013nfa}.
The data require the existence of a fast cooling process in the Cas A NS interior.
On the other hand, the NS cooling model must also
explain the compact object  XMMU J173203.3-344518 \cite{Klochkov}
in a supernova remnant, for which the surface temperature has recently
been measured.
This object is hotter and older than Cas A, at an age between 10 and 40 kyr.
Moreover, there exists information on surface temperatures of many other
NS sources.
It is not easy to appropriately explain these essentially different surface
temperatures of various objects within the so called ``minimal cooling
paradigm", where the only relevant rapid process is the so called
pair-breaking-formation (PBF) process on $3P_2$ paired neutrons.
The solution of the puzzle might be associated with a strong
medium dependence of cooling inputs, as provided by the
density (NS mass) dependent medium modifications of the nucleon-nucleon
interaction caused by the softening of the pion exchange contribution with increase of
the density, and by the density dependent superfluid pairing gaps,
see \cite{Migdal:1990vm,Voskresensky:2001fd} for details.
The key idea formulated long ago \cite{Voskresensky:1986af} is that the
cooling of various sources should be essentially different due to
the difference in their masses.
At that time there was the opinion that all NS masses should be fixed closely to
the value of $1.4 M_{\odot}$.
The recent measurements of the masses of the pulsars PSR J1614-2230 \cite{Demorest},
PSR J0348-0432 \cite{Anotoniadis} and J00737-3039B \cite{Kramer} and of the
companion of J1756-2251 \cite{Faulkner} have provided the proof
for the existence of NS masses varying in a broad range (at least from 1.2 to 2 $M_\odot$).

The most efficient processes within the ``nuclear medium cooling"
scenario are the medium modified Urca (MMU) processes, e.g.,  $nn\to
npe\bar{\nu}$ \cite{Voskresensky:1986af,Senatorov:1987aa}, and the
PBF processes $N\to
N_{pair}\nu\bar{\nu}$, $N=n$ or $p$
\cite{Flowers:1976ux,Voskresensky:1987hm}.
While being enhanced owing to their one-nucleon nature
\cite{Voskresensky:1987hm,Kolomeitsev:2008mc},
the latter processes are allowed only in
the presence of nucleon pairing.
Based on the assumption that the mass distribution of  those objects
for which surface temperatures are measured is similar to the one extracted, e.g.,
from a population synthesis, the very efficient direct Urca reaction,
$n\to npe\bar{\nu}$, should be forbidden in the majority of the former
NSs, see \cite{Klahn:2006ir,Blaschke:2006gd}.
The influence of in-medium effects on the NS cooling was first demonstrated
in \cite{Schaab:1996gd} with various  EoS.
The nuclear medium cooling scenario which was systematically developed
further in \cite{Blaschke:2004vq,Grigorian:2005fn,Blaschke:2011gc}
provides a successful description of the whole set of known cooling data for NSs with
low magnetic fields.
An overall fit of the data is obtained in our model for a strongly suppressed value
of the $3P_2$ neutron pairing gap, thus being in favour of  results by
\cite{Schwenk:2003bc}.

Reference \cite{Blaschke:2011gc}
has demonstrated an appropriate fit of the Cas A cooling curves with
results from ACIS-S instrument yielding a surface temperature decline of
$3\dots 4\%$ over 10 yrs.
For that description the lepton heat conductivity has been suppressed artificially by a
factor $\sim 0.3$ in the most favourable case compared to the result
of Ref.~\cite{Baiko:2001cj}, which has been exploited in our previous works. The nucleon contribution to the heat conductivity is suppressed in our case by medium effects compared to that used in Ref.~\cite{Baiko:2001cj}.
In the more recent work \cite{Blaschke:2013vma} we have
used the result of Ref.~\cite{Shternin:2007ee} for the lepton heat
conductivity which includes Landau damping effects. We have demonstrated that an appropriate fit of  the Cas A data was then possible
without applying any artificial suppression of the lepton heat conductivity.
The HHJ  EoS which has been exploited in our previous works has then been stiffened  in
\cite{Blaschke:2013vma} for $n \gsim 4n_0$
to comply with the constraint that the EoS
should allow for a maximum NS mass above the value $M=2.01\pm 0.04~M_\odot$ measured for
PSR J0348+0432 by \cite{Anotoniadis}, see also \cite{Demorest}.
However, the resulting EoS (labeled as HDD) which produces
$M_{\rm max}=2.06 M_{\odot}$  might still not be sufficiently stiff, since
the existence of even more massive objects than
those known by now \cite{Demorest,Anotoniadis} is not excluded.
Incorporating systematic light-curve differences the
authors of  \cite{Romani} have estimated that the mass of the
black-widow pulsar PSR J1311-3430 should at least be $M>2.1 M_{\odot}$.
Furthermore, a deconfinement transition in the NS interior
would contradict these NS mass measurements,  if one would use a  soft
hadronic EoS, cf. \cite{Alford:2006vz,Klahn,Klahn2}.
Therefore, the investigation of the possibility of hybrid stars requires a stiff
hadronic EoS.
Also, recent radius determinations from timing
residuals suggest larger radii (albeit still with large uncertainties) and
thus motivate the usage of a stiffer hadronic EoS
\cite{Bogdanov:2012md}.

Note that the authors of the recent paper \cite{Ho:2014pta}
have explained the Cas A ACIS-S data for the NS mass $M=1.44 M_{\odot}$
within the minimal cooling scenario using the BSk21 EoS,  a large proton gap
and a moderate $3P_2$ neutron gap.
Hottest and coldest objects, however, can hardly be explained appropriately
within the same scenario.
Note also for completeness that the authors of Ref.~\cite{Posselt:2013xva}
have argued that  the decline extracted from the ACIS-S  graded mode data
might be too steep.
In spite of this uncertainty in the analysis of the data we, as well as Ref.~\cite{Posselt:2013xva}, focus below mostly 
on a comparison with the ACIS-S graded mode data for Cas A.

 In the previous contribution \cite{Grigorian:2015nva} we have
demonstrated preliminary calculations of the cooling curves within our nuclear medium cooling
scenario exploiting the stiff DD2 EoS \cite{Typel:2009sy}.
However, in that work we have not performed any  tuning of the parameters of
the hadronic model. Additionally, as an alternative to the purely hadronic scenario, we
have incorporated in \cite{Grigorian:2015nva} the possibility of a deconfinement phase transition
from such a stiff nuclear matter EoS in the outer core to
color superconducting quark matter in the inner core.
Here we continue our study  within the hadronic scenario exploiting the stiff DD2 EoS
by tuning the proton pairing gaps and the effective
pion gap in order to construct a better  fit of the cooling data.
The deconfinement phase transition will be considered elsewhere.

\section{\label{sec:EoS}EoS of hadronic matter}

In our previous works \cite{Blaschke:2004vq,Grigorian:2005fn,Blaschke:2011gc} we
have exploited the HHJ ($\delta =0.2$) fit \cite{Heiselberg:1999fe} of
the APR EoS \cite{Akmal:1998cf}.
While the latter EoS produces an appropriate maximum NS mass of $M=2.2~M_{\odot}$,
the HHJ EoS  fit introduces an additional parametric correction of the high-density
behaviour in order to avoid a causality breach.
This comes at the price of a lowering of the maximum NS mass to $M=1.94~M_{\odot}$,
below the values for the measured masses of the pulsars PSR J1614-2230 \cite{Demorest}
and PSR J0348-0432 \cite{Anotoniadis}.
Recently we have modified this EoS by invoking an excluded volume for nucleons
\cite{Blaschke:2013vma}.
The so constructed HDD EoS stiffens for higher baryon density $n$ resulting in
an increase of the maximum NS mass up to the value $M_{max}=2.06~M_{\odot}$.

However, a recent radius determination for the nearest
millisecond pulsar PSR J0437-4715 \cite{Bogdanov:2012md}
indicates that a still stiffer EoS might be needed to support (at
$2\sigma$ confidence) radii  $\geq 13$ km  in the mass segment
between 1.5 and 1.8 $M_\odot$.
The density dependent relativistic mean-field EoS of Ref.~\cite{Typel:1999yq} with the well
calibrated DD2 parametrization \cite{Typel:2009sy} meets this requirement.
Moreover, the DD2 EoS fulfils  standard constraints for
symmetric nuclear matter around the saturation density and from nuclear structure.
The density dependent symmetry energy  agrees
with the  constraint by Danielewicz and Lee \cite{Danielewicz:2013upa} and with ab-initio
calculations for pure neutron matter \cite{Hebeler:2010jx}.
The direct Urca reaction threshold is not reached within the DD2 EoS.
However, due to the stiffness of DD2 EoS, it does not fulfil the  "flow
constraint" \cite{Danielewicz:2002pu} for densities above $2n_0$.
This is the price to be paid for the possibility to  increase the
maximum mass and the radii of NS.

 \begin{figure}[!thb]
   \centering
   \includegraphics[width=0.9\textwidth,keepaspectratio=true,angle=0]{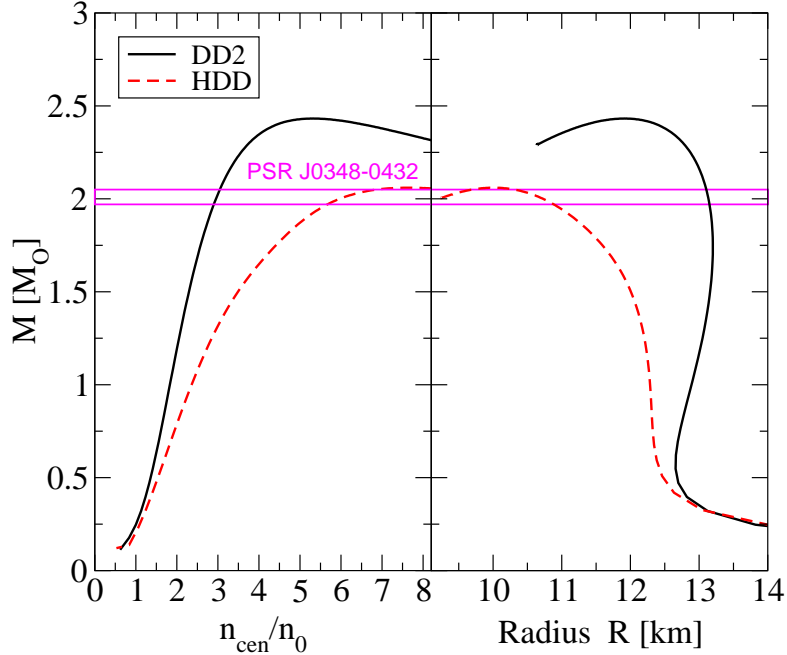}
     \caption{Mass vs. central baryon density (left) and vs. radius
     (right) for the stiff DD2  hadronic EoS (dash-dotted lines)
     and for the HDD EoS (solid lines). Crust is not included.}
   \label{Fig:MR}
 \end{figure}

The dependences of NS masses on central baryon density $n_{\rm cen}$  are shown in the left panel of
Fig.~\ref{Fig:MR} for the HDD and DD2 hadronic EoS.
We see that for a fixed NS mass the stiffening of the EoS leads to a redistribution of the density profile
in the NS interior so that the central densities get lowered.
As a consequence, a slower cooling is expected for stars with the DD2 EoS when compared with stars of
the same mass described by the HDD EoS.
In other words, in order to cover the same set of cooling data with a stiffer EoS the range of
masses attributed to the set of cooling curves shall be shifted to higher values.
The right panel of Fig.~\ref{Fig:MR} demonstrates the mass-radius relation (crust is not included).
The stiffer DD2 EoS produces a larger radius than the softer HDD EoS.

We shall now discuss the results for NS cooling obtained with both, the HDD and the DD2 EoS.

 \section{\label{sec:Cooling}Cooling }

In our previous works
\cite{Blaschke:2004vq,Grigorian:2005fn,Blaschke:2011gc,Blaschke:2013vma}
we have demonstrated that the cooling history is sensitive to the
efficiency of the medium modified Urca process controlled by the
density dependence of the effective pion gap shown in Fig.~1 of
\cite{Blaschke:2004vq} and to the value and the density dependence
of the $1S_0$ $pp$ pairing gap.
The results are very sensitive to the $3P_2$ $nn$ gap.
In our works we follow the analysis of \cite{Schwenk:2003bc} where this gap turns out
to be negligibly small.
Our results are rather insensitive to the treatment of the $1S_0$
neutron-neutron ($nn$) pairing gap since it is not spread deeply
to the interior region. Thereby this gap is taken the same as in our previous works, cf. \cite{Blaschke:2004vq}.  
In \cite{Blaschke:2011gc,Blaschke:2013vma} we have also demonstrated that the
decline of the cooling curve describing the evolution of the Cas A surface temperature
is sensitive to the value of the heat conductivity.
In \cite{Blaschke:2013vma} we demonstrated that at an appropriate choice of the proton pairing gap (following model I) we are able to fit  the 4$\%$ decline ACIS-S data on Cas A using
the same lepton heat conductivity as in \cite{Shternin:2007ee}. With the gaps given by model II
using
the same lepton heat conductivity we match  2$\%$ decline.

Here, we adopt the cooling inputs such as the neutrino emissivities, specific heat,
crust properties, etc., from our earlier works performed on the basis of the HHJ EoS
\cite{Blaschke:2011gc} and the HDD EoS \cite{Blaschke:2013vma} for hadronic matter.
The heat conductivity is the same as in \cite{Blaschke:2013vma}.
The best fit of Cas A data with the HDD EoS was obtained in \cite{Blaschke:2013vma}
with the same effective pion gap and the same $1S_0$ $pp$ pairing gap of the model I,
as in our previous works \cite{Blaschke:2004vq,Grigorian:2005fn,Blaschke:2011gc}.
Now exploiting the DD2 EoS to get the best fit of the cooling data we
will additionally tune the $pp$ pairing gap and the pion effective
gap retaining all other values the same as in \cite{Blaschke:2013vma}.

 \begin{figure}[!thb]
   \centering
   \includegraphics[width=0.8\textwidth,keepaspectratio=true,angle=0]{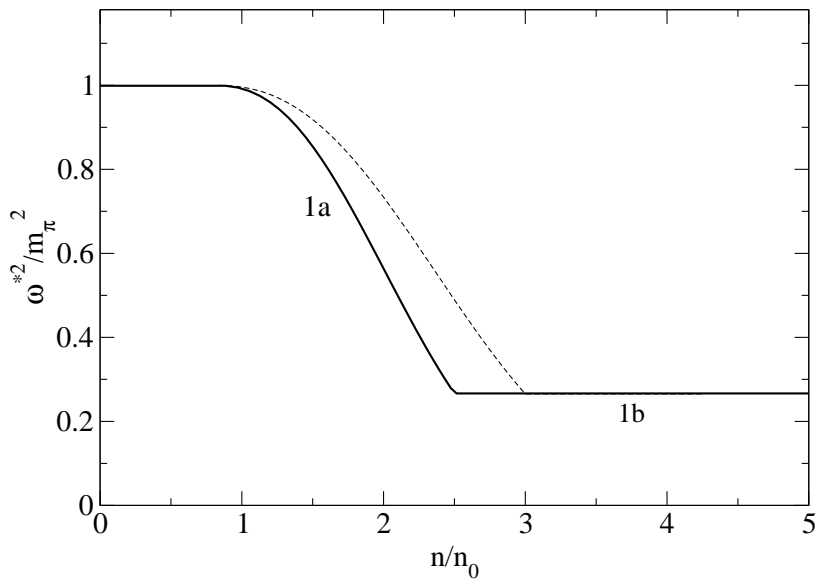}
     \caption{Square of the effective pion gap as a function of the density without pion
     condensation (curves 1a+1b), $m_{\pi}$ is the pion mass. The dotted line corresponds to
     the same parameterization as in our previous works, the solid line demonstrates a stronger
     softening effect.}
   \label{Fig:omegatilde}
 \end{figure}

The density dependence of the square of the effective pion gap ${\omega}^{*\,2} (n)$
that we exploit in the given work is shown in Fig.~\ref{Fig:omegatilde}.
To be specific we consider the case when the pion softening is saturated and pion condensation
does not occur. 
The dotted curve 1a+1b is precisely the same as in Fig.~1 of \cite{Blaschke:2004vq}, demonstrating saturation of the pion softening for $n>3n_0$.
The solid line shows a stronger pion softening effect with a saturation for $n>2.5 n_0$. 

 \begin{figure}[!thb]
   \centering
   \includegraphics[width=0.9\textwidth,angle=0]{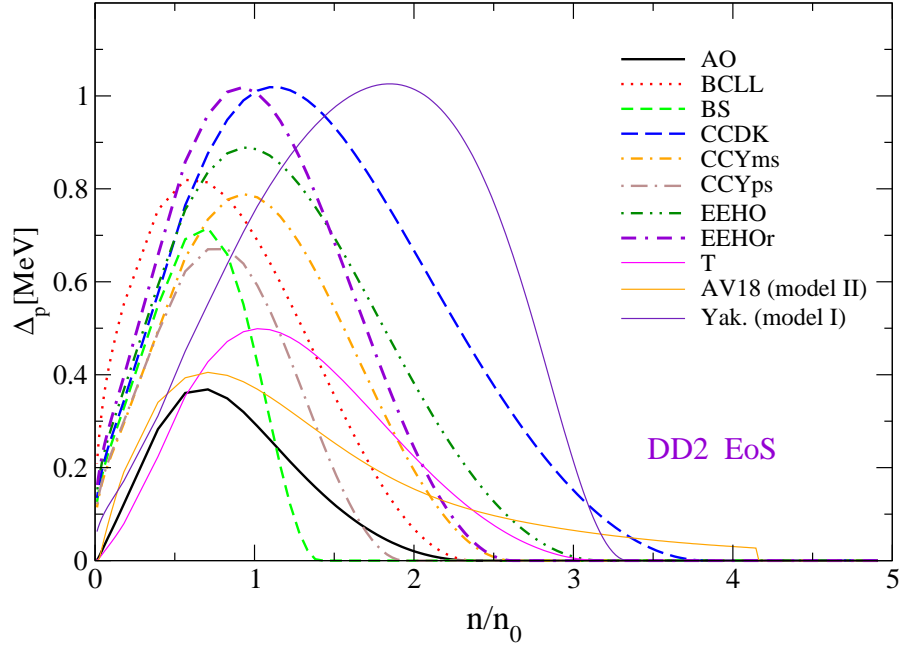}
     \caption{$1S_0$ $pp$ gaps as  functions of the proton density.
     The abbreviations in the legend correspond to those used in Ref.~\cite{Ho:2014pta}.
     The gaps labeled as ``Yak" and ``AV18" are those (models I and II, respectively)
     exploited in our previous works
     \cite{Blaschke:2004vq,Grigorian:2005fn,Blaschke:2011gc,Blaschke:2013vma}. }
   \label{Fig:gaps}
 \end{figure}

The $1S_0$ $pp$ pairing gaps exploited by different authors are
shown in Fig.~\ref{Fig:gaps} for the DD2 EoS.
We use the parametrization of the $pp$ pairing gaps, $\Delta_p (p_{{\rm
F},p})$, from \cite{Ho:2014pta}, Eq. (2),  $p_{{\rm F},i}$ denotes the
Fermi momentum of the species $i$.
The parameters are taken to fit the gaps computed in various publications.
The abbreviations of the curves in Fig.~\ref{Fig:gaps} are taken over from Table II of \cite{Ho:2014pta}.
Two additional curves labeled as ``Yak" and ``AV18" correspond to the models I and II, respectively, exploited in our previous works \cite{Blaschke:2004vq,Grigorian:2005fn,Blaschke:2011gc} for the HHJ
EoS and in \cite{Blaschke:2013vma} for the HDD EoS.

With these gaps and the $\omega^*$ parametrizations we compute the NS cooling history.

\begin{figure}[!thb]
   \centering
   \includegraphics[width=0.9\textwidth,angle=0]{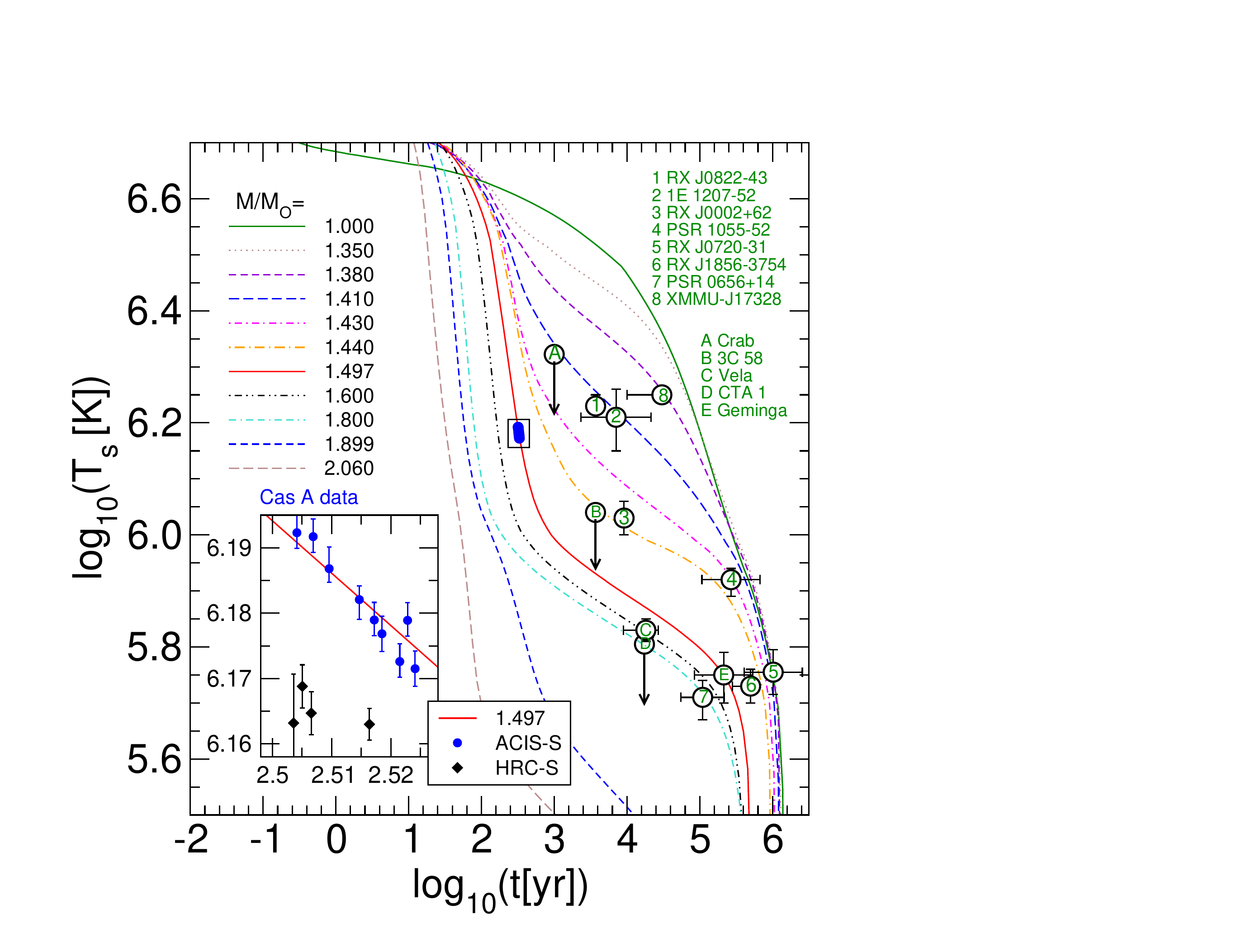}
      \caption{Cooling curves for a NS sequence according to the  hadronic HDD EoS;
      $T_s$ is the redshifted surface temperature, $t$ is the NS age.
      The effective pion gap is given by the dotted curve 1a+1b in Fig.~\ref{Fig:omegatilde}.
      The $1S_0$ $pp$ pairing gap corresponds to model I.
      The  mass range is shown in the legend. Cooling data for Cas A are
       explained with a NS of mass $M=1.497~M_\odot$.
       A comparison with Cas A ACIS-S and HRC-S data is shown in the inset. }
   \label{Fig:Cool1}
 \end{figure}

\section{\label{sec:Results}Results}

In Fig.~\ref{Fig:Cool1} we demonstrate the NS cooling history
computed with the HDD EoS in  Ref.~\cite{Blaschke:2013vma} using
model I for the $pp$ pairing gaps (see the gap ``Yak" in Fig.~\ref{Fig:gaps})
and with the effective pion gap given by the dotted 1a+1b line  in Fig.~\ref{Fig:omegatilde}.
As we see, the Cas A data are described by a NS with the mass $M_{\rm CasA} \simeq
1.497 M_{\odot}$.
A slight change of the value $M_{\rm CasA}$ compared to the value $1.54 M_{\odot}$ found in
\cite{Blaschke:2013vma} is due to inessential modifications of the parametrization in the present work.

The stiffer DD2 hadronic EoS  compared to those for the
HHJ and HDD EoS produces a smaller central density for the star of the given mass
and therefore it leads to a weaker cooling activity, provided the same inputs are used for the
effective pion gap ${\omega}^{*} (n)$, the pairing gaps and
the other model ingredients. As the result, a  larger NS mass  is required 
to describe Cas A  cooling with the stiffer EoS.
Ref.~\cite{Grigorian:2015nva} demonstrated  that when changing the EoS 
a description of all cooling data is possible even without changing any of
the formerly adjusted cooling inputs except a tuning of the heat conductivity
(in line with the strategy applied before in Ref.~\cite{Blaschke:2011gc}), but
the NS mass that is required to fit the Cas A cooling data with very stiff DD2 EoS
then amounts to $M=2.426~M_\odot$.

 \begin{figure}[!thb]
   \centering
   \includegraphics[width=0.9\textwidth,angle=0]{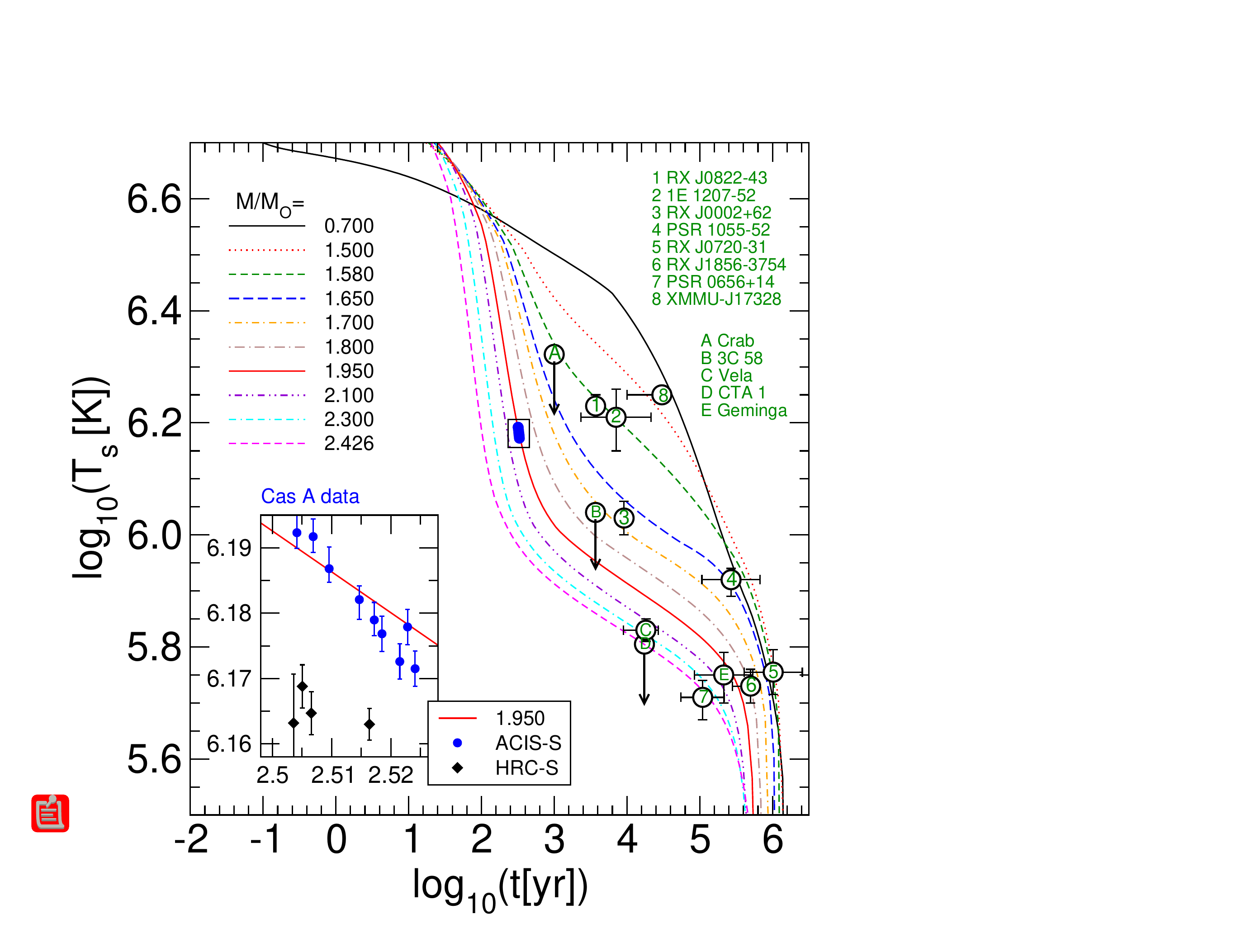}
      \caption{
      Same as Fig.~\ref{Fig:Cool1} but for the hadronic DD2 EoS and the $1S_0$ $pp$ 
      pairing gap corresponding to the model EEHOr.
      The cooling data for Cas A are now explained with a NS of mass $M=1.950~M_\odot$. }
   \label{Fig:Cool2}
 \end{figure}

Now exploiting DD2 EoS we take the heat conductivity the same as in \cite{Blaschke:2013vma}
(without any additional tuning) but tune the effective pion gap and the $pp$
pairing gap.
This allows to describe Cas A cooling data by a NS with a lower mass. We performed calculations with all the gap curves shown in
Fig.~\ref{Fig:gaps} and with $\omega^* (n)$ given by the dotted
and solid curves 1a+1b (without pion condensation) in Fig.~\ref{Fig:omegatilde}.

The resulting cooling curves are shown in Fig.~\ref{Fig:Cool2}
for the $pp$ pairing gap of the model EEHOr and for the effective pion gap given by the dotted curve 1a+1b.
With other $pp$ pairing gaps we get a less regular description of the cooling data.
The cooling of the hot source XMMU-J1732 is explained by a NS with
the mass $\sim 1.5~ M_{\odot}$.
At rather low densities relevant for stars with a mass $M\sim M_{\odot}$
the pion softening effect is not pronounced.
A large $pp$ proton gap at such densities is required, otherwise the cooling curves
go down. The description of the coldest objects requires a $pp$ gap dropping to zero at
central densities reached in those massive objects. The cooling is
determined by the efficient medium modified Urca process (in the absence of pairing).
The description of the steep decline of the cooling curve for Cas A as indicated by the
ACIS-S requires a small heat conductivity (in our case the lepton heat conductivity is computed
following \cite{Shternin:2007ee} and the nucleon contribution incorporates a
decrease with increasing density owing to the pion softening effect)
and an efficient medium modified Urca process. Cooling data for Cas A are
       explained with a NS of $M=1.950~M_\odot$.

   \begin{figure}[!htb]
   \centering
   \includegraphics[width=0.9\textwidth,angle=0]{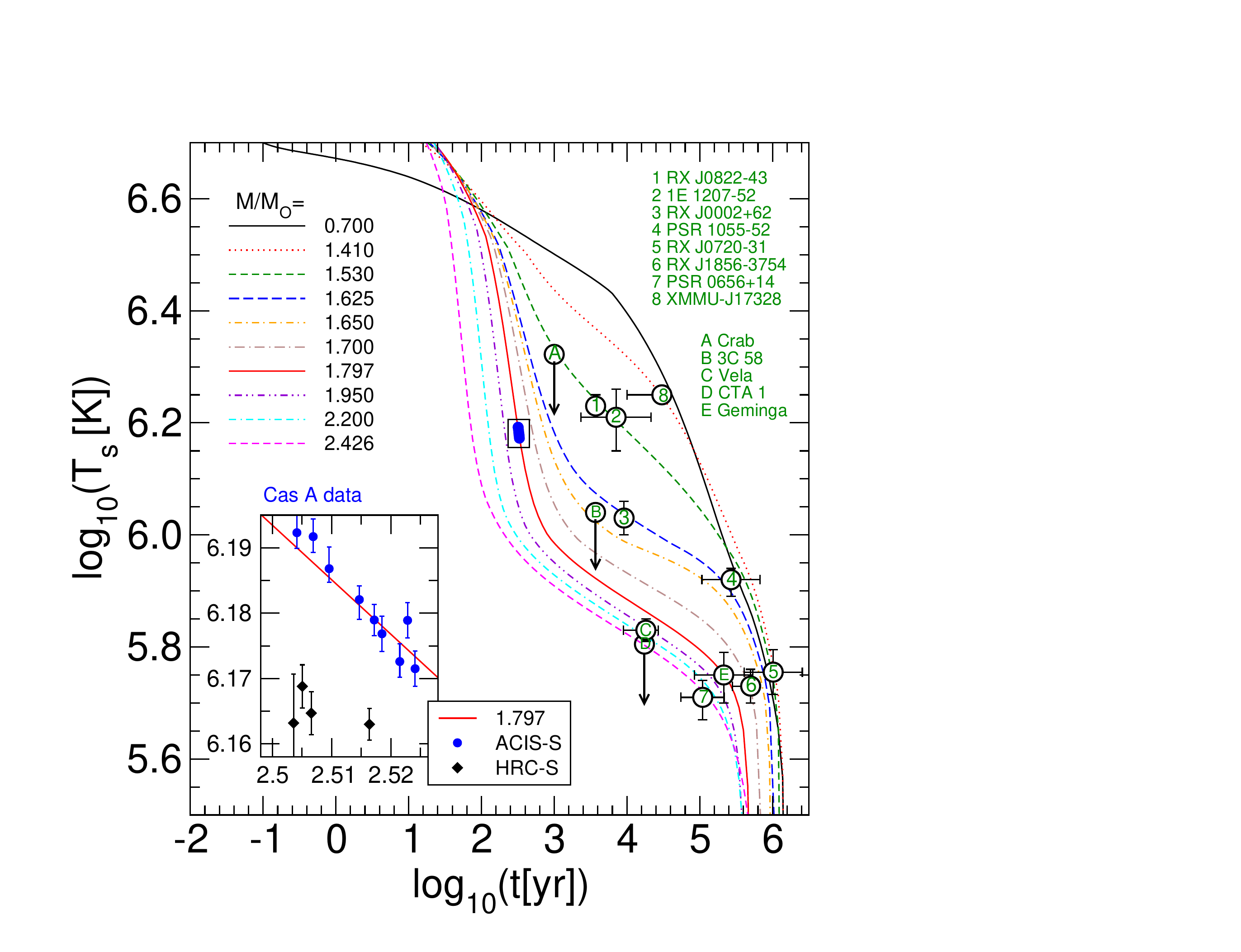}
      \caption{
      Same as Fig.~\ref{Fig:Cool2}, but with the
      effective pion gap given by the solid curve 1a+1b in Fig.~\ref{Fig:omegatilde}. 
      The cooling data for Cas A are now explained with a NS of mass $M=1.797~M_\odot$.}
   \label{Fig:Cool3}
 \end{figure}

Fig.~\ref{Fig:Cool3} demonstrates the same as  Fig.~\ref{Fig:Cool2}
 but with the effective pion gap calculated with the help of the
 solid curve 1a+1b in Fig.~\ref{Fig:omegatilde}.
 We see that the Cas A cooling data can now be explained with a lower NS mass, 
 of  $M=1.797~M_\odot$. The cooling of the hot source XMMU-J1732 is also explained by a NS with a smaller mass, $\sim 1.4~ M_{\odot}$. These values could  be still decreased if we assumed a steeper $\omega^* (n)$ dependence than that we have chosen. Note that variational calculations of 
 Ref.~\cite{Akmal:1998cf}
 show that pion condensation in neutron star matter may appear already for $n\simeq 1.3 n_0$, been in favor of a steeper $\omega^* (n)$ dependence. Just in order to be conservative as much as possible we continue to exploit a much weaker pion softening in our calculations.

  \begin{figure}[!htb]
   \centering
   \includegraphics[width=0.9\textwidth,angle=0]{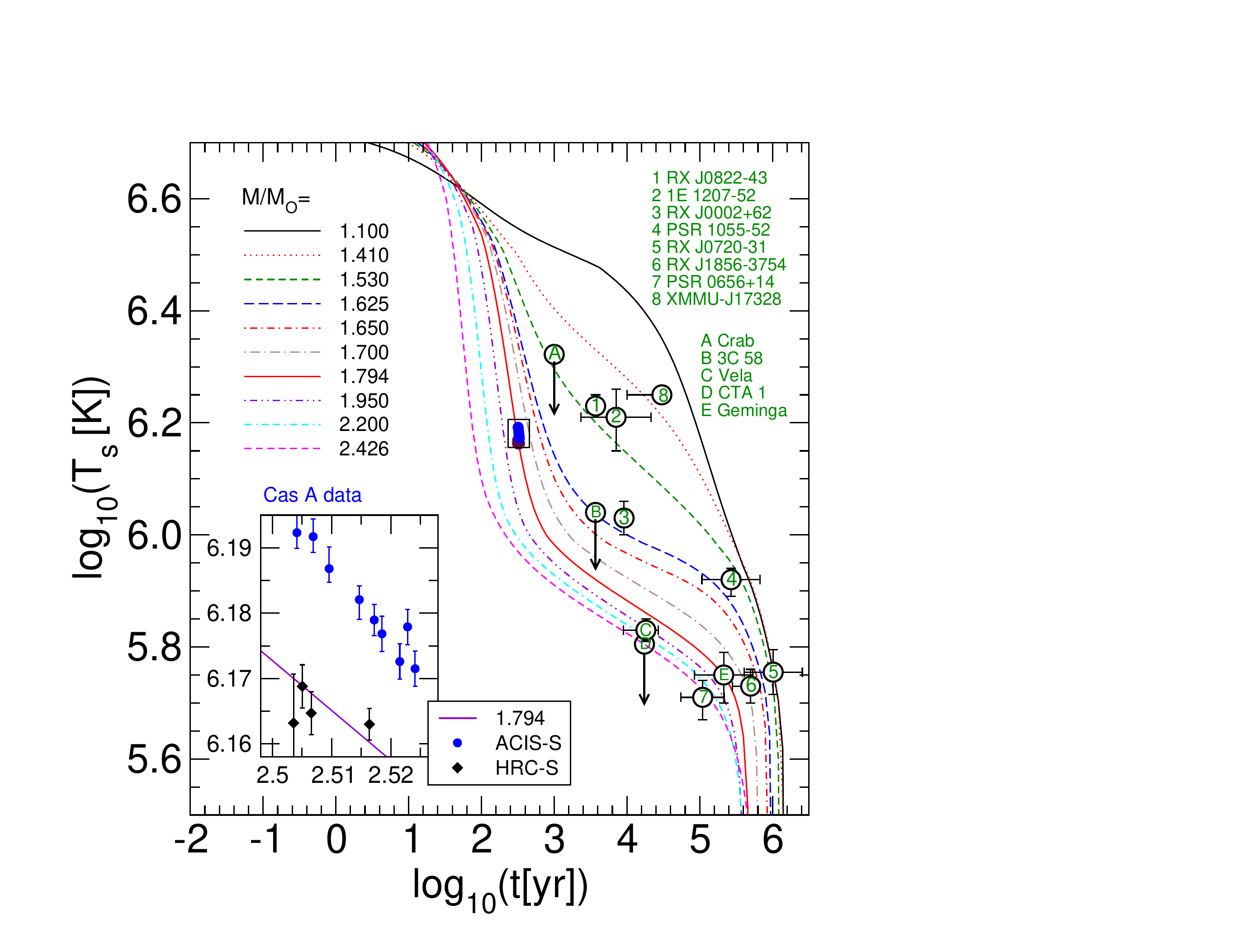}
      \caption{
      Same as Fig.~\ref{Fig:Cool3}, but with the $1S_0$ $pp$ pairing gap corresponding to the 
      model CCYms.
      The cooling data for Cas A are now explained with a NS of mass $M=1.794~M_\odot$.  }
   \label{Fig:Cool4}
 \end{figure}

In Fig.~\ref{Fig:Cool4} we show the same as in Fig. \ref{Fig:Cool3} but for the  $1S_0$ $pp$ pairing gaps corresponding to model CCYms. With these gaps we reproduce HRC-S data for the NS mass $M=1.794~M_\odot$, which is only slightly different from $M=1.797~M_\odot$, with which we  reproduced ACIS-S data.

\section{\label{sec:Remarks}Concluding remarks}

As we demonstrated  with the DD2 EoS, large
values of the NS radii and the maximum mass, as they might be
motivated by observations, are compatible with our nuclear medium
cooling scenario provided one uses a stiff EoS.
The presently known cooling data on Cassiopeia A and the hot source XMMU-J1732,
as well as other cooling data, are appropriately described  by
purely hadronic stars within the nuclear medium cooling scenario, under the
assumption that different sources have different masses.
The resulting cooling curves are sensitive to the value and the
density dependence of the $pp$ pairing gap and the effective pion gap.
Choosing the $pp$ pairing gap such that it is large for
densities relevant to low mass objects and disappearing for higher
densities met in centres of more massive stars we are able to
reach an overall agreement with the cooling data including
Cassiopeia A and the hot source XMMU-J1732, exploiting soft as well as
stiff hadronic EoS.
Exploiting a stiff EoS and allowing the effective pion gap to decrease with
increasing density we are able to diminish the value of the mass of the neutron star
in Cassiopeia A required for an optimal description of its cooling data.
Fitting a steep decline of the cooling curve for Cassiopeia A requires a rather low value of the heat
conductivity, an appropriate form of the $pp$ gap and an efficient
medium modified Urca neutrino emissivity. 

\subsection*{Acknowledgments}

This work was supported by the project 4807/PB/IFT/15,
UMO-2014/13/B/ST9/02621
and by  the Ministry of Education and Science of the Russian
Federation (Basic part). H.G. acknowledges support by the
Bogoliubov-Infeld programme  for exchange between JINR Dubna and
Polish Institutes. The authors are grateful for support from the
COST Action MP1304 "NewCompStar" for their networking and
collaboration activities.
The work of D.B. has been supported by
the Hessian LOEWE initiative through HIC for FAIR.

\end{document}